\documentclass[12pt]{article}

\usepackage{graphicx}
\usepackage{slashed}
\begin{document}

\title{Proton Decay and the Origin of Quark and Lepton Mixing}
\author{{\bf S.M. Barr} and {\bf Heng-Yu Chen} \\
Department of Physics and Astronomy \\ 
and \\
Bartol Research Institute \\ University of Delaware
Newark, Delaware 19716} \maketitle

\begin{abstract}
It was recently proposed that all flavor mixing has a single source,
namely the mixing of the three quark and lepton families with ``extra" vectorlike 
fermions in ${\bf 5} + \overline{{\bf 5}}$ multiplets of $SU(5)$. 
This was shown to lead to several testable predictions including 
neutrino masses and CP-violating phases. Here it is shown that
the mixing angles within grand unified fermion multiplets are also
predicted. Proton decay branching ratios would thus give several
independent tests of the model. Certain model parameters could be 
determined independently from the quark and lepton spectrum and
from proton decay. 
\end{abstract}

\section{Introduction} 

In a recent paper \cite{BarrChen}, we proposed a highly predictive model
that relates the MNS mixing of the leptons \cite{MNS} and the CKM mixing 
of the quarks \cite{CKM} by positing a single source for all flavor
violation. The purpose of the present paper is to point out that the same model
also predicts all the mixing angles that come into gauge-boson-mediated
proton decay, thus giving further tests of the model. 

The basic idea of the model, which is based on $SU(5)$, is that all 
flavor violation
comes from the mixing of the ``usual" three families of fermion multiplets, which we shall denote  
${\bf 10}^U_I + \overline{{\bf 5}}^U_I$, $I=1,2,3$, with
some ``extra" vectorlike multiplets, which we shall denote ${\bf 5}^E_A +\overline{{\bf 5}}^E_A$, 
$A=1,2,...,N$. In particular, it is assumed that the Yukawa terms involving 
only the ``usual" fermions are flavor-diagonal due to an abelian family symmetry. 
The ``extra" vectorlike fermions, on the other hand, do not transform under this 
symmetry, and as a consequence their Yukawa couplings to the usual fermions produce
flavor violation. What 
ends up happening is that a single matrix, describing the 
mixing of the usual $\overline{{\bf 5}}^U$ multiplets with the extra 
$\overline{{\b 5}}^E$ multiplets, controls all flavor violation. It turns out,
moreover,  
that this matrix, which we call $A_{\Delta}$, can be completely determined from the quark masses 
and CKM parameters alone, allowing several predictions of neutrino properties, 
as well as placing
non-trivial constraints on the values of several other quantities that are as yet
not very precisely determined. 

Just as the CKM and MNS angles specify how fermion mass eigenstates
are arranged within the multiplets of the electroweak $SU(2)_L$ group,
there are mixing angles that specify how the fermions are arranged within 
the multiplets of the grand unification group. Since the model we are discussing posits
a single source for all flavor mixing, it predicts these grand unified mixing angles 
also. This means that a measurement of proton decay branching ratios would give 
non-trivial tests of the model. We extract the predictions for proton decay
coming from the exchange of the grand unified gauge bosons, which would dominate
in a non-supersymmetric version of the model. (The model is easily supersymmetrized. 
That would not significantly affect
the predictions for quark and lepton properties found in \cite{BarrChen}. It would,
however, presumably mean that the dominant proton decay amplitudes would come from
exchange of colored Higgsinos. We do not discuss those here.) 

What makes this model so predictive is the combination of $SU(5)$ symmetry, which relates the quarks and leptons, and the abelian flavor symmetry, which forbids flavor violation in the sector of the ``usual" fermion multiplets. One of the interesting features of the model is that even though it unifies quarks and leptons, it explains in a simple and natural way why the CKM mixing angles of the former are much smaller than the MNS mixing angles of the latter. 
The point is that ultimate source of all flavor violation is the mixing among $\overline{{\bf 5}}$
multiplets, which contain the left-handed {\it leptons}, but not the left-handed quarks.
This is the basic idea that underlies so-called ``lopsided models" \cite{lopsided}, of
which the present model is a particularly simple and predictive example.

This model is also very closely related to a model proposed in 1984 as a solution to the Strong CP Problem \cite{NelsonBarr}. The mechanism for solving the Strong CP Problem proposed there can therefore be easily implemented in the present model. If it is, then
further predictions result, though they would require measuring proton decay branching ratios to about a percent accuracy. That the mechanism can be tested at all, however, is quite striking and surprising, as the relevant physics happens at the GUT scale. Again, what makes that
possible here is the powerful combination of $SU(5)$ symmetry and abelian flavor symmetry.
(Incidentally, the specific abelian symmetry does not really matter, as long as it renders those Yukawa terms that involve only the ``usual" fermions flavor diagonal.)

In section 2, we review the model and how it gives predictions for neutrino properties and other quantities. We go into more detail here in our treatment of complex phases than we did in \cite{BarrChen} as this is necessary to determine the relative phases of certain proton-decay amplitudes. The simple basic forms predicted by the model for the various quark and lepton mass matrices are shown in Eq. (6). These result in a prediction of the entire $3 \times 3$ complex mass matrix of the neutrinos (which contains 9 real observables) in terms of two complex parameters and one real parameters. This prediction is displayed in Eq. (14).
In section 3, we derive the predictions for the $SU(5)$ mixing angles and 
proton decay branching ratios. In section 4, we see that by embedding the model in $SO(10)$, which is very simply done, even more interesting proton decay predictions result. The most interesting proton decay predictions are shown in
Eqs. (26) - (28).

\section{Review of the model} 

We shall now briefly review the details of the model before proceeding to show
how it predicts the angles that come into proton decay. More details can be found at \cite{BarrChen}. The Yukawa terms of the model are 

\begin{equation}
\begin{array}{ccl}
{\cal L}_{Yuk} & = & Y_{_I} ({\bf 10}^U_I {\bf 10}^U_I) \langle {\bf 5}_h
\rangle + y_{_I} ({\bf 10}^U_I \overline{{\bf 5}}^U_I) \langle {\bf
5}^{\dag}_h \rangle
\\ & & \\
& + & \tilde{Y}_{_I} ({\bf 10}^U_I {\bf 10}^U_I) \langle {\bf 45}_h \rangle
+ \tilde{y}_{_I} ({\bf 10}^U_I \overline{{\bf 5}}^U_I) \langle {\bf
45}^{\dag}_h \rangle
\\ & & \\
& + & (\lambda_{_I}/M_R)(\overline{{\bf 5}}^U_I \overline{{\bf
5}}^U_I) \langle {\bf 5}_h \rangle \langle {\bf 5}_h \rangle
\\ & & \\
& + &  Y'_{_{AB}} ({\bf 5}^E_A \overline{{\bf 5}}^E_B) \langle {\bf 1}_h
\rangle + y'_{_{AI}} ({\bf 5}^E_A \overline{{\bf 5}}^U_I) \langle {\bf
1}'_{hI} \rangle,
\end{array}
\end{equation}

\noindent where the subscript $h$ denotes Higgs multiplets. Note that the last two terms give $SU(5)$-invariant masses to fermions in the ${\bf 5}$ and $\overline{{\bf 5}}$
multiplets, so that these masses can be much larger than the weak scale and possibly
even the GUT scale. It does not matter to the analysis in this paper what the scale of these $SU(5)$-invariant masses are, so long as the last two terms in Eq. (1) are roughly
the {\it same} scale. We will only assume that these masses are much heavier than the weak scale, and refer to them as ``heavy"; while any fermions whose masses are less than or at the weak scale (i.e. the Standard Model fermions) we will call ``light". There are many 
abelian symmetries that could enforce the flavor-diagonal form of the terms
in the first three lines of Eq. (1). A simple example is that given in 
\cite{BarrChen}, namely $K_1 \times K_2 \times K_3 \times K'$, where (for a
given $I$ equal to 1,2, or 3) $K_I$ is a $Z_2$ symmetry under which
${\bf 10}^U_I$, $\overline{{\bf 5}}^U_I$, and ${\bf 1}'_{hI}$ are odd
and all other fields even. $K'$ is a $Z_N$ symmetry ($N>2$) under
which ${\bf 5}^E_A \rightarrow e^{2\pi i/N} {\bf 5}^E_A$,
$\overline{{\bf 5}}^E_A \rightarrow e^{2 \pi i/N} \overline{{\bf
5}}^E_A$, ${\bf 1}_h \rightarrow e^{-4 \pi i/N} {\bf 1}_h$, and 
${\bf 1}'_{hI} \rightarrow e^{-2 \pi i/N} {\bf 1}'_{hI}$. One of the roles played by
the $K'$ symmetry is to forbid the Higgs fields that break $SU(5)$ at the superlarge
scales (for example, adjoint Higgs fields) from coupling to the fermions.

The Yukawa terms in Eq. (1) that involve only the usual fermion multiplets
give rise to mass matrices $m_u$, $m_d$, $m_{\ell}$, and $m_{\nu}$
that are flavor diagonal:

\begin{equation} 
u_{_{10^U_I}} \; (m_u)_{_{IJ}} \; u^c_{_{10^U_J}},  \;\;\;\;\;
d_{_{10^U_I}} \; (m_d)_{_{IJ}} \; d^c_{_{\overline{5}^U_J}}, \;\;\;\;\;
\ell_{_{\overline{5}^U_I}} \; (m_{\ell})_{_{IJ}} \; \ell^c_{_{10^U_J}}, \;\;\;\;\;
\nu_{_{\overline{5}^U_I}} \; (m_{\nu})_{_{IJ}} \; \nu_{_{\overline{5}^U_J}}
\end{equation}

\noindent However, the fermions in $\overline{{\bf 5}}^U_I$, $I = 1,2,3$, are not simply the observed Standard Model fermions. The $\overline{{\bf 5}}^U_I$ and the $\overline{{\bf 5}}^E_A$ mix, with some linear combinations of them becoming ``heavy", and the three combinations orthogonal to them remaining ``light". We denote these respectively by $\overline{{\bf 5}}^H$ and $\overline{{\bf 5}}^L$. To identify these heavy and light multiplets one must look at the heavy fermion masses. From the last two terms in Eq. (1), these are seen to have the form

\begin{equation}
{\bf 5}^E_A \left( [y'_{_{AI}} \langle {\bf 1}_{hI} \rangle] \; \overline{{\bf 5}}^U_I + [Y'_{_{AB}} \langle {\bf 1}_h \rangle] \; \overline{{\bf 5}}^E_B
\right) \;\; \equiv \;\; 
{\bf 5}^E_A  \underbrace{(\Delta_{_{AI}} \overline{{\bf 5}}^U_I + M_{_{AB}} \overline{{\bf 5}}^E_B)}_{\propto \;\; \overline{{\bf 5}}^H_A},
\end{equation} 
 
\noindent One sees that $\overline{{\bf 5}}^H \propto (\Delta \overline{{\bf 5}}^U + M \overline{{\bf 5}}^E)$. One can easily write the light $\overline{{\bf 5}}^L$ as the linear combinations orthogonal to these and then invert to write the ``usual" multiplets 
$\overline{{\bf 5}}^U$ as linear combinations of $\overline{{\bf 5}}^L$ and $\overline{{\bf 5}}^H$. The result turns out to be (suppressing indices)

\begin{equation}
\begin{array}{rcl}
\overline{{\bf 5}}^U & = & A \overline{{\bf 5}}^L + B 
\overline{{\bf 5}}^H, \\ & & \\
{\rm where} \;\;\; A & \equiv & [I + \Delta^{\dag} M^{-1 \dag} M^{-1} \Delta]^{-1/2}.
\end{array} 
\end{equation}

\noindent where $A$ and $B$ are complex matrices. ($A$ is $3 \times 3$ and $B$ is $3 \times N$, where $N$ is the number of ``extra" $\overline{{\bf 5}}$ multiplets.) For the ${\bf 10}$ multiplets, there are no heavy mass terms or mixing with ``extra" multiplets, so the ``usual" multiplets are simply the same as the light multiplets: ${\bf 10}^U = {\bf 10}^L$. If we rewrite Eq. (2) in terms of the light fermion multiplets using Eq. (4), and suppress indices for clarity, 
we obtain 

\begin{equation} 
u^T_{_{10^L}} \; m_u \; u^c_{_{10^L}},  \;\;\;\;\;
d^T_{_{10^L}} \; m_d \; A \; d^c_{_{\overline{5}^L}}, \;\;\;\;\;
\ell^T_{_{\overline{5}^L}} \; A^T \; m_{\ell} \; \ell^c_{_{10^L}}, \;\;\;\;\;
\nu^T_{_{\overline{5}^L}} \; A^T \; m_{\nu} \; A \; \nu_{_{\overline{5}^L}}
\end{equation}

\noindent Note that the matrix $A$ appears next to the fermions that are in $\overline{{\bf 5}}$ multiplets, because $A$ reflects the effects of the mixing of those multiplets. The terms in Eq. (5) give the effective mass matrices of the three families of Standard Model fermions, which we will denote by {\it capital} $M$: $M_u$, $M_d$, $M_{\ell}$, and $M_{\nu}$. One sees from Eq. (5) that these matrices
are products of the {\it diagonal} matrices $m_u$, $m_d$, $m_{\ell}$, and $m_{\nu}$
in Eq. (2) and the matrix $A$, as follows:

\begin{equation}
M_u = m_u, \;\;\;\;\;\; M_d = m_d A, \;\;\;\;\;\; M_{\ell} = A^T m_{\ell}, \;\;\;\;\;\;
M_{\nu} = A^T m_{\nu} A.
\end{equation}

We thus see that all flavor violation is controlled by $A$. Moreover, the matrix $A$ can be brought to a simple form in the following way. By multiplying $A$ on the right by a unitary matrix, the elements below the main diagonal of $A$ can be made zero. Then by rescaling the rows by multiplying from the left by a complex diagonal matrix, the diagonal elements of $A$ can be set to 1. Thus, $A$ can be written 

\begin{equation}
A = {\cal D} \; A_{\Delta} \; {\cal U},
\end{equation}

\noindent where ${\cal D}$ is a complex diagonal matrix, ${\cal U}$ is a unitary matrix, and
$A_{\Delta}$ is a matrix of the form

\begin{equation}
A_{\Delta} = \left( \begin{array}{ccc} 1 & b & c e^{i \theta} \\ 0 & 1 & a \\ 0 & 0 & 1 \end{array}
\right),
\end{equation}

\noindent where $a$, $b$, and $c$ are real. ($a$ and $b$ have been made real by absorbing phases in ${\cal D}$ and ${\cal U}$.) Since it will be important to keep track of complex phases for the later analysis of proton decay, let us define ${\cal D} \equiv e^{i \Phi} D$, where $D$ and $\Phi$ are real and diagonal matrices, and similarly $m_u \equiv e^{i \Phi_u} m^R_u$, 
$m_d \equiv e^{i \Phi_d} m^R_d$, $m_{\ell} \equiv e^{i \Phi_{\ell}} m^R_{\ell}$, $m_{\nu} \equiv e^{i \Phi_{\nu}} m^R_{\nu}$, where $\Phi_u$, $m^R_u$, etc. are real and diagonal matrices. 

It will be convenient to define a ``flavor basis" of $SU(5)$ multiplets (denoted by superscript $F$) by $\overline{{\bf 5}}^F \equiv
{\cal U} \overline{{\bf 5}}^L$ and ${\bf 10}^F \equiv e^{i (\Phi + \Phi_d)} {\bf 10}^L$. (There are three families of these, but we are suppressing the family indices.)  Therefore, if we use Eq. (7) to rewrite the expressions in Eq. (5), we can absorb the factors of ${\cal U}$ and some of the phases into redefined fermion multiplets, as follows.

\begin{equation} 
\begin{array}{cl}
u^T_{_{10^L}} \; m_u \; u^c_{_{10^L}} & =
u^T_{_{10^L}} \; (e^{i \Phi_u} m^R_u) \; u^c_{_{10^L}} =
u^T_{_{10^F}} \; m^R_u \; e^{i(\Phi_u -2 \Phi_d - 2 \Phi)} u^c_{_{10^F}} \\ & \\
& = u^T_{_{10^F}} \; \overline{m}_u \; e^{i \Theta_{u^c}} u^c_{_{10^F}}, \;\;\;\;\; 
\;\;\;\; {\rm where} \;\; \overline{m}_u \equiv m^R_u, \;\; \Theta_{u^c} \equiv \Phi_u - 2 \Phi_d - 2 \Phi,
\\ & \\ 
d^T_{_{10^L}} \; (m_d \; A) \; d^c_{_{\overline{5}^L}} & =
d^T_{_{10^L}} \; (e^{i \Phi_d} m^R_d) \; (e^{i \Phi} D A_{\Delta} {\cal U}) \; d^c_{_{\overline{5}^L}} = 
d^T_{_{10^F}} \; (m^R_d D) \; A_{\Delta} \; d^c_{_{\overline{5}^F}}
\\ & \\
& = d^T_{_{10^F}} \; (\overline{m}_d  \; A_{\Delta}) \; d^c_{_{\overline{5}^F}}, \;\;\;\;\;
\;\;\;\; {\rm where} \;\; \overline{m}_d \equiv m^R_d D,  
\\ & \\
\ell^T_{_{\overline{5}^L}} \; (A^T \; m_{\ell}) \; \ell^c_{_{10^L}} 
& = \ell^T_{_{\overline{5}^L}} \; ({\cal U}^T  A^T_{\Delta} D e^{i \Phi}) 
\; (e^{i \Phi_{\ell}} m^R_{\ell}) \; \ell^c_{_{10^L}} 
= \ell^T_{_{\overline{5}^F}} \; A^T_{\Delta} \; (D \overline{m}^R_{\ell}) 
\; e^{i (\Phi_{\ell} - \Phi_d)} \; \ell^c_{_{10^F}} 
\\ & \\
& = \ell^T_{_{\overline{5}^F}} \; (A^T_{\Delta} \overline{m}_{\ell}) 
\; e^{i \Theta_{\ell^c}} \; \ell^c_{_{10^F}}, \;\;\;\;\;
\;\;\;\; {\rm where} \;\; \overline{m}_{\ell} \equiv D m^R_{\ell}, 
\;\; \Theta_{\ell^c} \equiv \Phi_{\ell} - \Phi_d, 
 \\ & \\ 
\nu^T_{_{\overline{5}^L}} \; (A^T \; m_{\nu} \; A) \; \nu_{_{\overline{5}^L}}
& = 
\nu^T_{_{\overline{5}^L}} {\cal U}^T \; A^T_{\Delta} \; (e^{2i (\Phi_{\nu} + \Phi)}
D^2 m^R_{\nu}) \; A_{\Delta} \; {\cal U} \; \nu_{_{\overline{5}^L}} 
= \nu^T_{_{\overline{5}^F}} \; A^T_{\Delta} \; e^{2i (\Phi_{\nu} + \Phi)} 
(D^2 m^R_{\nu}) \; A_{\Delta} \; \nu_{_{\overline{5}^F}} \\ & \\ 
& =  
\nu^T_{_{\overline{5}^F}} \; (A^T_{\Delta} \; e^{i \Theta_{\nu}} \overline{m}_{\nu} \; A_{\Delta}) \; \nu_{_{\overline{5}^F}}, \;\;\;\;\;\;\;\;\; 
\overline{m}_{\nu} \equiv D^2 m^R_{\nu}, \;\; \Theta_{\nu} \equiv \Phi_{\nu} + \Phi.
\end{array} 
\end{equation}

\noindent This gives the mass matrices of the three Standard Model families in the ${\bf 10}^F$, $\overline{{\bf 5}}^F$ basis as

\begin{equation}
\overline{M}_u = \overline{m}_u e^{i \Theta_{u^c}}, \;\;\;\;\;\; \overline{M}_d = \overline{m}_d A_{\Delta}, \;\;\;\;\; \overline{M}_{\ell} = A_{\Delta}^T \overline{m}_{\ell} e^{i \Theta_{\ell^c}}, \;\;\;\;\;
\overline{M}_{\nu} = A_{\Delta}^T (e^{i \Theta_{\nu}} \overline{m}_{\nu}) A_{\Delta}.
\end{equation}

\noindent Note that so far no transformations have been done that contribute to CKM or MNS mixing. To get to the ${\bf 10}^F, \overline{{\bf 5}}^F$ basis from the ${\bf 10}^L, \overline{{\bf 5}}^L$ basis, we have done a transformation by ${\cal U}$ to the $\overline{{\bf 5}}$ multiplets, i.e. to (i) the left-handed anti-down quarks (right-handed quarks), which does not affect CKM mixing, and (ii) the left-handed charged leptons and neutrinos, which, because it was the {\it same} for the charged leptons and neutrinos, does not affect MNS mixing.   

The parameters that come into the quark and lepton masses and the CKM and MNS mixing
have been reduced to the matrix $A_{\Delta}$; the four real diagonal
matrices $\overline{m}_u$, $\overline{m}_d$, $\overline{m}_{\ell}$, and $\overline{m}_{\nu}$; and two relative phases in $\Theta_{\nu}$. 

Now consider the diagonalization of the mass matrices in Eq. (10), which must be done to find the physical quarks and leptons, i.e. the mass eigenstates, which shall be denoted with a tilde. The matrix $\overline{M}_u$ is already diagonal. It can be made real by redefining the phases of the anti-up quarks. So define $\tilde{u}  = u_{10^F}$ and $\tilde{u}^c = e^{i \Theta_{u^c}} u_{10^F}$. The matrix
$\overline{M}_d = \overline{m}_d \; A_{\Delta}$ has the form

\begin{equation}
\overline{M}_d = \overline{m}_d \; A_{\Delta} = \overline{\mu}_d \left(
\begin{array}{ccc} \overline{\delta}_d & 0 & 0 \\
0 & \overline{\epsilon}_d & 0 \\ 0 & 0 & 1 \end{array} \right) \; 
\left( \begin{array}{ccc} 1 & b & c e^{i \theta} \\ 0 & 1 & a \\ 0 & 0 & 1 \end{array}
\right) = \overline{\mu}_d \left(
\begin{array}{ccc} \overline{\delta}_d & \overline{\delta}_d b & \overline{\delta}_d c e^{-i \theta} \\
0 & \overline{\epsilon}_d & \overline{\epsilon}_d a \\ 0 & 0 & 1 \end{array} \right),
\end{equation} 

\noindent where we use the same names for the parameters as in \cite{BarrChen}. The fact that $m_d \ll m_s \ll m_b$ tells us that $\overline{\delta}_d \ll \overline{\epsilon}_d \ll 1$.
From this it is easy to see that $\overline{M}_d$ is diagonalized by a unitary rotation from the left (to set to zero the elements above the diagonal) and a negligible rotation from the right. In fact, since $\overline{M}_u$ is already diagonal, one sees that the transformation from the left needed to diagonalize $\overline{M}_d$ is just the CKM matrix: 

\begin{equation}
\overline{M}_d = \overline{m}_d \; A_{\Delta} \cong V^*_{_{CKM}} \; m_d^{DIAG} = V^*_{_{CKM}} \left( \begin{array}{ccc}
m_d & 0 & 0 \\ 0 & m_s & 0 \\ 0 & 0 & m_b \end{array} \right).
\end{equation}

\noindent The mass eigenstates are therefore 
$\tilde{d} = V_{_{CKM}}^{\dag} d_{10^F} \; \; \Rightarrow \; \; d_{10^F} = V_{_{CKM}} \tilde{d}$, whereas for the anti-down quarks, one has simply $\tilde{d}^c \cong d^c_{\overline{5}^F}$.
From Eqs. (11) and (12) it is easy to determine the matrix $A_{\Delta}$:

\begin{equation} 
a = \frac{m_b}{m_s} \; V_{cb} \approx 2, \;\;\;\;\;\; b = \frac{m_s}{m_d} \; \overline{V}_{us} \approx 4, \;\;\;\;\;\;
ce^{i \theta} = \frac{m_b}{m_s} |V_{cb}| e^{i \delta_{_{CKM}}} \approx 3 e^{i \delta_{_{CKM}}}.
\end{equation}

\noindent Here $\overline{V}_{us} \equiv \tan \theta_C$, where 
$V_{us} \equiv \sin \theta_C$. (For the other CKM angles the difference between sine and tangent is numerically insignificant.) So the matrix $A_{\Delta}$ is completely determined by measured quark parameters.

The diagonalization of $\overline{M}_{\ell}$ proceeds in an analogous way, except transposed. So the non-negligible rotation in this case is done from the right, i.e. to the left-handed anti-leptons $\ell^c$, whereas a negligible transformation is needed of the left-handed leptons $\ell$. We can write 
$A^T_{\Delta} \; \overline{m}_{\ell} = m_{\ell}^{DIAG} \; V^{\dag}_{\ell^c}$. Therefore the mass eigenstates of the anti-leptons are $\tilde{\ell}^c = V_{\ell^c}^{\dag} 
e^{i \Theta_{\ell^c}} \ell^c_{10^F}$, $\tilde{\ell} \cong \ell_{\overline{5}^F}$. Since negligible rotation of the left-handed charged leptons is needed to diagonalize $\overline{M}_{\ell}$, the MNS mixing comes almost entirely from the diagonalization of the neutrino mass matrix $\overline{M}_{\nu}$, which is given by

\begin{equation}
\begin{array}{ccl}
\overline{M}_{\nu} & = & A^T_{\Delta} (e^{i \Theta_{\nu}} \overline{m}_{\nu}) A_{\Delta} \\ & & \\
&\cong & \left[ \begin{array}{ccc} 1 & 0 & 0 \\
\frac{m_s}{m_d} | \overline{V}_{us}| & 1 & 0 \\
\frac{m_b}{m_d} |V_{ub}| e^{i \delta} & \frac{m_b}{m_s} |V_{cb}| & 1
\end{array} \right]
\left[ \begin{array}{ccc} q e^{i \beta} & 0 & 0
\\ 0 & p e^{i \alpha} & 0 \\
0 & 0 & 1 \end{array} \right] \left[
\begin{array}{ccc}
1 & \frac{m_s}{m_d} |\overline{V}_{us}| &
\frac{m_b}{m_d} |V_{ub}| e^{i \delta} \\
0 & 1 & \frac{m_b}{m_s} |V_{cb}| \\
0 & 0 & 1 \end{array} \right] \mu_{\nu},
\end{array}
\end{equation}

\noindent where we have pulled out an overall factor $\mu_{\nu}$ to scale the 
33 element of $\overline{m}_{\nu}$ to 1. Since the transformation to diagonalize $\overline{M}_{\nu}$ is to a very good approximation just the MNS matrix, one has
$\tilde{\nu} = U^{\dag}_{_{MNS}} \; \nu_{\overline{5}^F} \;\; \Rightarrow \;\; \nu_{\overline{5}^F} = U_{_{MNS}} \tilde{\nu}$.

Eq. (14) is the key equation as far as the prediction of neutrino properties is concerned. It contains five real parameters $p, \alpha, q, \beta, \mu_{\nu}$ to fit nine neutrino parameters that are (in principle) observable, namely 
three neutrino masses, three MNS angles, and three CP phases (one ``Dirac" and two ``Majorana"). There are consequently four predictions. In addition, as explained in \cite{BarrChen}, there are strong constraints on parameters that have already been measured but with large errors bars. For example the fits obtained in \cite{BarrChen} favor values of $\theta_{atm} \leq 45^{\circ}$, values of $\theta_{solar} \leq 34^{\circ}$, values of $m_s/m_d \leq 20$, and values of the CKM phase $\leq  1.27$ radians. 

We can now write the content of the $SU(5)$ multiplets that contain the Standard Model fermions, in the ${\bf 10}^F$, $\overline{{\bf 5}}^F$ basis. Using the facts that
$u_{10^F} = \tilde{u}, \;\;\; u^c_{10^F} = \tilde{u}^c, \;\;\; d_{10^F} = V_{_{CKM}} \; \tilde{d}, \;\;\;
d^c_{\overline{5}^F} = \tilde{d}^c, \;\;\; \ell^c_{10^F} = V_{\ell^c} \; \tilde{\ell}^c, \;\;\; \ell_{\overline{5}^F} = \tilde{\ell}$, and $\nu_{\overline{5}^F} = U_{_{MNS}} \tilde{\nu}$, one has 

\begin{equation} 
{\bf 10}^F_I = \left( [e^{-i\Theta_{\ell^c}} V_{\ell^c} \tilde{\ell}^c], \;\; \left( \begin{array}{c} \tilde{u}
\\ V_{_{CKM}} \tilde{d} \end{array} \right), \;\; [e^{-i \Theta_{u^c}} \tilde{u}^c] \right)_I, \;\;\;\;\; 
\overline{{\bf 5}}^F_I = \left( \begin{array}{c} \left[ \begin{array}{c} U_{_{MNS}} \tilde{\nu} \\ \tilde{\ell} \end{array} \right] \\ \tilde{d}^c \end{array} \right)_I,
\end{equation} 

\noindent where $I$ is the family index. So, for example, if $I = 1$, we see that the $SU(5)$ partners of the physical $u_L$ are (i) the physical $u^c_L$ times a phase, (ii) the CKM linear combination of the physical $d_L$, $s_L$, and $b_L$, namely $(V_{ud} d_L + V_{us} s_L + V_{ub} b_L)$, and (iii) the linear combination of the physical anti-leptons $((V_{\ell^c})_{11} e^c + (V_{\ell^c})_{12} \mu^c + (V_{\ell^c})_{13} \tau^c)$ times a phase. 

\section{Proton decay angles for $SU(5)$ modes}

Let us consider proton decay caused by the exchange of the superheavy gauge
bosons of $SU(5)$. These obviously only make transitions within the irreducible multiplets of $SU(5)$. From Eq. (15), therefore, we see that the only mixing matrices that enter into 
such proton decay amplitudes are $U_{_{MNS}}$, $V_{_{CKM}}$, and $V_{\ell^c}$. The CKM and MNS matrices can be measured at low-energy, and are fairly well known. The matrix $V_{\ell^c}$ cannot be measured at low energy, but is predicted by the model, since it comes from diagonalizing $A^T_{\Delta} \overline{m}_{\ell}$. $A_{\Delta}$ is known (and given in Eqs. (8) and (13)). The diagonal matrix $\overline{m}_{\ell}$ can be determined from the masses of the charged leptons. One finds that to an excellent approximation

\begin{equation}
V_{\ell^c} =  \left( \begin{array}{ccc} 1 & -\frac{m_e}{m_{\mu}} b & -\frac{m_e}{m_{\tau}} c e^{i \theta} \\ \frac{m_e}{m_{\mu}} b & 1 & - \frac{m_{\mu}}{m_{\tau}} a \\
\frac{m_e}{m_{\tau}} c e^{-i \theta} & \frac{m_{\mu}}{m_{\tau}} a & 1
\end{array} \right) \approx \left( \begin{array}{ccc}
1 & -0.02 & -0.001 e^{i \theta} \\ 0.02 & 1 & -0.12 \\
0.001 e^{-i \theta} & 0.12 & 1 \end{array}
\right),
\end{equation}
  
\noindent Since the $\tau$ lepton is too heavy to be involved in proton decay, only the 12  element of $V_{\ell^c}$ , which we shall call $\zeta$, enters the proton decay amplitudes. 
($\zeta \cong 0.02$.) 

As can be seen from Eq. (15), this model gives quite definite predictions for all the mixing angles that come into $SU(5)$ fermion multiplets and thus into the amplitudes for proton decay via the exchange of the superheavy $SU(5)$ gauge bosons. In particular, one sees that the only mixings that come into those amplitudes are the CKM and MNS elements and the (small) 12 element of $V_{\ell^c}$. 

While this is a definite and (in principle) testable prediction, it is not very distinctive of this model, as all models will involve the CKM and MNS mixing. The one non-zero parameter that comes into proton decay that is distinctive of this model is $\zeta$. Because $\zeta$ is small, however, it would be very hard to measure even if proton decay is seen. On the other hand, if $\zeta$ {\it were} measured, then, as we will show, it would allow a test of a well-known mechanism \cite{NelsonBarr} for solving the Strong CP Problem that is otherwise almost impossible to test.  

If we embed this model in $SO(10)$, as is quite natural and simple to do, one finds predictions for proton decay branching ratios that are quite distinctive of this model. In fact, they would allow an independent measurement of the parameter $b$ and a combination of $a$ and $c$, which are also determined by low-energy physics (see Eqs. (8) and (13)). We shall consider the $SO(10)$ embedding of the model and the resulting proton decay predictions in the next section. First, we shall find the proton-decay operators coming from the exchange of the $SU(5)$ gauge bosons.

Let us denote the superheavy gauge bosons of $SU(5)$ by $W^a_i$, where $a = 3,4,5$ is an $SU(3)_c$ color index and $i = 1,2$ is an $SU(2)_L$ weak index. The relevant couplings 
are $(\overline{\psi})_{ac} \slashed{W}^a_i \psi^{ic} + (\overline{\psi})_{aj} \slashed{W}^a_i \psi^{ij} + (\overline{\psi})^i \slashed{W}^a_i \psi_a$, which in Standard Model notation, dropping the color and weak indices, and putting in the family indices $I$ are  $\overline{u^c}_I \slashed{W} Q_I + \overline{Q}_I \slashed{W} \ell^c_I + \overline{L}_I \slashed{W} d^c_I$. Here the family indices refer to the ${\bf 10}^F$, $\overline{{\bf 5}}^F$ basis, shown in Eq. (15). These terms give the $d=6$ operators

\begin{equation}
\begin{array}{l}
\frac{g^2_5}{2 M^2_5} \; \left( \overline{Q}_I \gamma_{\mu} u^c_I \right) \left( \overline{L}_J \gamma^{\mu} d^c_J \right) - 
\frac{g^2_5}{2 M^2_5} \; \left( \overline{Q}_I \gamma_{\mu} u^c_I \right) \left( \overline{Q}_J \gamma^{\mu} \ell^c_J \right)\\ \\
= \frac{g^2_5}{2 M^2_5} \; \left[ \left( \overline{d}_I \gamma_{\mu} u^c_I \right) \left( \overline{\nu}_J \gamma^{\mu} d^c_J \right) + \left( \overline{u}_I \gamma_{\mu} u^c_I \right) \left( \overline{\ell}_J \gamma^{\mu} d^c_J \right) \right] \\ \\
- \frac{g^2_5}{2 M^2_5} \; \left[ \left( \overline{u}_I \gamma_{\mu} u^c_I \right) \left( \overline{d}_J \gamma^{\mu} \ell^c_J \right) + \left( \overline{d}_I \gamma_{\mu} u^c_I \right) \left( \overline{u}_J \gamma^{\mu} \ell^c_J \right) \right]. 
\end{array}
\end{equation}

\noindent Referring to Eq. (15), and keeping only the fermions light enough to be decay products of a nucleon, the relevant operators are

\begin{equation}
\begin{array}{l}
\frac{g^2_5}{2 M^2_5} \; \left( \overline{[\cos \theta_C \tilde{d} + \sin \theta_C \tilde{s}]} \gamma_{\mu} \tilde{u}^c \right) \left( \overline{\nu}_e \gamma^{\mu} \tilde{d}^c 
+ \overline{\nu}_{\mu} \gamma^{\mu} \tilde{s}^c \right) \\ \\
+ \frac{g^2_5}{2 M^2_5} \;  \left( \overline{\tilde{u}} \gamma_{\mu} \tilde{u}^c \right) \left( \overline{\tilde{e}} \gamma^{\mu} \tilde{d}^c  +
\overline{\tilde{\mu}} \gamma^{\mu} \tilde{s}^c \right) \\ \\
- \frac{g^2_5}{2 M^2_5} \;  \left( \overline{\tilde{u}} \gamma_{\mu} \tilde{u}^c \right) \left( \overline{[\cos \theta_C \tilde{d} + \sin \theta_C \tilde{s}]} \gamma^{\mu} [\tilde{e}^c + e^{-i\alpha_1} \zeta \tilde{\mu}^c] \right) \\ \\ 
- \frac{g^2_5}{2 M^2_5} \;  \left( \overline{\tilde{u}} \gamma_{\mu} \tilde{u}^c \right) \left( \overline{[- \sin \theta_C \tilde{d} + \cos \theta_C \tilde{s}]} \gamma^{\mu} [- e^{-i\alpha_2} \zeta \tilde{e}^c + \tilde{\mu}^c] \right) 
\\ \\
- \frac{g^2_5}{2 M^2_5} \; \left( \overline{[\cos \theta_C \tilde{d} + \sin \theta_C \tilde{s}]} \gamma_{\mu} \tilde{u}^c \right) \left( \overline{\tilde{u}} \gamma^{\mu} 
[\tilde{e}^c + e^{-\alpha_1} \zeta \tilde{\mu}^c] \right),
\end{array}
\end{equation}

\noindent where $\alpha_1$ and $\alpha_2$ are the 11 and 22 elements of the diagonal matrix $\Theta_{\ell^c}$.  Collecting like terms (and doing a Fierz transformation of the last term in Eq. (18)) we obtain:

\begin{equation}
\begin{array}{l}
\frac{g^2_5}{2 M^2_5} \; \left( \overline{[\cos \theta_C \tilde{d} + \sin \theta_C \tilde{s}]} \gamma_{\mu} \tilde{u}^c \right) \left( \overline{\nu}_e \gamma^{\mu} \tilde{d}^c 
+ \overline{\nu}_{\mu} \gamma^{\mu} \tilde{s} \right) \\ \\
+ \frac{g^2_5}{2 M^2_5} \;  \left( \overline{\tilde{u}} \gamma_{\mu} \tilde{u}^c \right) \left( \overline{\tilde{e}} \gamma^{\mu} \tilde{d}^c  +
\overline{\tilde{\mu}} \gamma^{\mu} \tilde{s}^c \right) \\ \\
- \frac{g^2_5}{2 M^2_5} \;  \left( \overline{\tilde{u}} \gamma_{\mu} \tilde{u}^c \right) \left[ 2 c_{\theta} (\overline{\tilde{d}} \gamma^{\mu} 
\tilde{e}^c) + (2 s_{\theta} - e^{-i \alpha_2} \zeta) (\overline{\tilde{s}} \gamma^{\mu} 
\tilde{e}^c) + (2 e^{-i \alpha_1} \zeta - s_{\theta}) (\overline{\tilde{d}} \gamma^{\mu} 
\tilde{\mu}^c) - c_{\theta} (\overline{\tilde{s}} \gamma^{\mu} \tilde{\mu}) \right], 
\end{array}
\end{equation}

\noindent where $s_{\theta} \equiv \sin \theta_C = V_{us}$, $c_{\theta} \equiv \cos \theta_C$, and where we have dropped terms of order $\zeta s_{\theta}$ ($\approx 0.004$) or smaller.

The phases $\alpha_1$ and $\alpha_2$ are unknown free parameters of the model. They enter, however, only in the small terms proportional to $\zeta$, and thus their effect would not be significant unless the proton decay branching ratios were measurable to better than a percent accuracy. If the $O(\zeta)$ terms could be measured precisely enough, however, it would allow a test of a well-known mechanism for solving the Strong CP Problem, as we will now explain. As noted in \cite{BarrChen}, the model we are describing here implements in a very simple way the mechanism for solving the Strong CP Problem proposed in \cite{NelsonBarr}. Indeed, it {\it is} the model proposed in Nelson's paper, except that here we have imposed a flavor symmetry to make the Yukawa terms involving the ``usual" fermion multiplets diagonal. To solve the Strong CP Problem (assuming no supersymmtry) all that is needed is to impose CP as an invariance of the Lagrangian that is spontaneously broken by the Higgs fields we denoted ${\bf 1}'_{hI}$ in Eq. (1).
What would happen in that case is that the phases matrices $\Phi_u$, $\Phi_d$, $\Phi_{\ell}$, and $\Phi_{\nu}$ would all vanish. (The phases matrix $\Phi$ would {\it not} vanish, however, as it comes from the matrix $A$ that arises ultimately from
$\langle {\bf 1}'_{hI} \rangle$.) One can see from Eq. (9), that $\Theta_{\ell^c}$ is given by 
$\Phi_{\ell} - \Phi_d$ and therefore would vanish. Thus the mechanism for solving the Strong CP Problem \cite{NelsonBarr} predicts that $\alpha_1 = \alpha_2 = 0$.

\section{Proton decay angles in the $SO(10)$ embedding of the model}

More interesting predictions for proton decay arise if the model is embedded in $SO(10)$. This embedding is very simple. The ``usual" fermion multiplets are contained in spinors of $SO(10)$, while the ``extra" vectorlike fermion multiplets are contained in vectors of $SO(10)$:  

\begin{equation}
{\bf 10}^U + \overline{{\bf 5}}^U \subset {\bf 16}^U, \;\;\;\;\;\;\;\; {\bf 5}^E + \overline{{\bf 5}}^E \subset {\bf 10}^E. 
\end{equation}

\noindent Then the Yukawa terms in Eq. (1) come from the following $SO(10)$-invariant terms:

\begin{equation}
\begin{array}{ccl}
{\cal L}_{Yuk} & = & Y_{_I} ({\bf 16}^U_I {\bf 16}^U_I) \langle {\bf 10}_h \rangle
\\ & & \\
& + & \tilde{Y}_{_I} ({\bf 16}^U_I {\bf 16}^U_I) \langle \overline{{\bf 126}}_h \rangle
\\ & & \\
& + & (\lambda_{_I}/M_R)({\bf 16}^U_I {\bf
16}^U_I) \langle \overline{{\bf 16}}_h \rangle \langle \overline{{\bf 16}}_h \rangle
\\ & & \\
& + &  Y'_{_{AB}} ({\bf 10}^E_A \overline{{\bf 10}}^E_B) \langle {\bf 1}_h
\rangle + y'_{_{AI}} ({\bf 10}^E_A {\bf 16}^U_I) \langle {\bf
16}_{hI} \rangle.
\end{array}
\end{equation}

\noindent The analysis presented in sections 2 and 3 goes through without change. 
Now, however, there are additional superheavy gauge bosons that mediate proton decay, namely those that make transitions between the ${\bf 10}^U$ and $\overline{{\bf 5}}^U$ within the spinors ${\bf 16}^U$.  These gauge bosons transform as $(3, 2, \frac{1}{6}) + conj.$ under the Standard Model gauge group $SU(3)_c \times SU(2)_L \times U(1)_Y$, and can be denoted by $W^{ai}$ and $W_{ai}$, where $a$ is a color index and $i$ is a weak index.
These gauge bosons have coupling to the ${\bf 10}^U$ and $\overline{{\bf 5}}^U$ in the ${\bf 16}^U$ of the form $\overline{\psi}_{\alpha \beta} \slashed{W}_{\gamma \delta}  
\psi_{\eta} \epsilon^{\alpha \beta \gamma \delta \eta}$ (where $\alpha$, $\beta$, etc. are $SU(5)$ indices), which contains
$(\overline{\psi}_{bc} \slashed{W}_{ai} \psi_j - 
\overline{\psi}_{bj} \slashed{W}_{ai} \psi_c) \epsilon_{ijabc}$. Translating this into the notation of the Standard Model, one has 

\begin{equation}
(\overline{u^c})^a \slashed{W}_{ai} (L)_j \epsilon^{ij} + (\overline{Q})_{jb}
\slashed{W}_{ai} (d^c)_c \epsilon^{abc} \epsilon^{ij}.
\end{equation} 

\noindent The fermions in these operators are in the ``usual" multiplets ${\bf 10}^U + \overline{{\bf 5}}^U \subset {\bf 16}^U$. Let us make this explicit, but suppress the $SU(5)$ indices for clarity:
$\overline{u^c}_{10^U} \slashed{W} L_{\overline{5}^U} + \overline{Q}_{10^U}
\slashed{W} d^c_{\overline{5}^U}$. Using Eqs. (4) and (7), this can be rewritten
in terms of the fields in the multiplets ${\bf 10}^F$ and $\overline{{\bf 5}}^F$ shown in Eq. (15):

\begin{equation}
\begin{array}{cl}
& \overline{u^c}_{10^U} \slashed{W} L_{\overline{5}^U} + \overline{Q}_{10^U}
\slashed{W} d^c_{\overline{5}^U} \\ & \\ 
\longrightarrow & \overline{u^c}_{10^L} \slashed{W} A L_{\overline{5}^L} + \overline{Q}_{10^L}
\slashed{W} A d^c_{\overline{5}^L} \\ & \\
= & \overline{u^c}_{10^L} \slashed{W} {\cal D} A_{\Delta} ({\cal U} L_{\overline{5}^L}) + \overline{Q}_{10^L}
\slashed{W} {\cal D} A_{\Delta} ({\cal U} d^c_{\overline{5}^L}) \\ & \\
= & e^{i ( \Phi_d + \Phi)} \left[ \overline{u^c}_{10^F} \slashed{W} [{\cal D} A_{\Delta}] L_{\overline{5}^F} + \overline{Q}_{10^F}
\slashed{W} [{\cal D} A_{\Delta}] d^c_{\overline{5}^F} \right]
\end{array}
\end{equation} 

The gauge bosons in Eq. (23) form a weak doublet with electric charges $\frac{1}{3}$ and $- \frac{2}{3}$, which we will denote $W_{(1/3)}$ and $W_{(-2/3)}$. First, let us consider the couplings of $W_{(1/3)}$, which will give the proton decay modes with charged leptons. Writing out the terms in the last line of Eq. (23), referring to Eq. (15), and keeping only those quarks and leptons that are lighter than a nucleon, one has 

\begin{equation}
\begin{array}{cl}
& e^{i(\Phi_u - \Phi_d)_1} [ \overline{\tilde{u}^c} \slashed{W}_{(1/3)} D_1 \sum_{J=1,2} (A_{\Delta})_{1J} \tilde{\ell}_J
+  \overline{\tilde{u}} \slashed{W}_{(1/3)} D_1 \sum_{J=1,2} (A_{\Delta})_{1J} \tilde{d}^c_J] \\ & \\
= & e^{i(\Phi_u - \Phi_d)_1} D_1 [\overline{\tilde{u}^c} \slashed{W}_{(1/3)}
(\tilde{e} + b \tilde{\mu})  
+ \overline{\tilde{u}} \slashed{W}_{(1/3)} (\tilde{d}^c + b \tilde{s}^c)], 
\end{array}
\end{equation}

\noindent where $D_J$ stands for the $JJ$ element of the diagonal matrix
$D$. The couplings in Eq. (24) give the $d=6$ nucleon decay operators

\begin{equation}
\begin{array}{cl}
& \left( \frac{g_{10}^2}{2M_{10}^2} (D_1)^2 \right) 
(\overline{[\tilde{e} + b \tilde{\mu}]} \gamma_{\mu} \tilde{u}^c ) \; ( \overline{\tilde{u}} \gamma^{\mu}[\tilde{d}^c + b \tilde{s}^c]) \\ & \\
=  & \kappa \left( \frac{g_{5}^2}{2M_{5}^2} \right) 
(\overline{\tilde{u}} \gamma_{\mu} \tilde{u}^c ) \; ( \overline{[\tilde{e} + b \tilde{\mu}]} \gamma^{\mu}[\tilde{d}^c + b \tilde{s}^c]),
\end{array}
\end{equation}

\noindent where the second line was obtained by a Fierz transformation, and we have defined the real parameter $ \kappa \equiv (D_1)^2 \frac{g_{10}^2/M_{10}^2}{g_5^2/M_5^2}$. 

Combining these $SO(10)$ operators with the $SU(5)$ operators given in Eq. (19), we can compute the proton-decay rates for the two-body decay modes having charged leptons in the final state, namely $p \rightarrow \pi^0 e^+$, $p \rightarrow \pi^0 \mu^+$, $p \rightarrow K^0 e^+$, and $p \rightarrow K^0 \mu^+$. If we consider the three ratios
of these four rates, most of the unknown quantities cancel out. Using the results of \cite{NathPerez} we obtain

\begin{equation}
\frac{\Gamma(p \rightarrow \pi^0 \mu^+)}{\Gamma(p \rightarrow \pi^0 e^+)}
= \frac{(b \kappa)^2 + |s_{\theta} - 2 e^{-i \alpha_1} \zeta |^2}{(1 + \kappa)^2 +
(2 c_{\theta})^2 }, 
\end{equation}

\begin{equation}
\frac{\Gamma(p \rightarrow K^0 e^+)}{\Gamma(p \rightarrow \pi^0 e^+)}
= R
\frac{(b \kappa)^2 + |2s_{\theta} - e^{-i \alpha_2} \zeta |^2}{(1 + \kappa)^2 +
(2 c_{\theta})^2 }, 
\end{equation}

\begin{equation}
\frac{\Gamma(p \rightarrow K^0 \mu^+)}{\Gamma(p \rightarrow \pi^0 e^+)}
= R
\frac{(1 + b^2 \kappa)^2 + c_{\theta}^2}{(1 + \kappa)^2 +
(2 c_{\theta})^2 },
\end{equation}

\noindent with

\begin{equation}
R \equiv 2 \left( 1 - \frac{m_K^2}{m_p^2} \right)^2 
\left( \frac{1 + \frac{m_p}{m_B} (D-F)}{1 + D + F} \right)^2 = 0.105 \pm 0.005
\end{equation}

\noindent where $D$ and $F$ are chiral lagrangian parameters found in \cite{CSW} 
to be $D+F = 1.267 \pm 0.003$ and $D-F = -0.341 \pm 0.016$ and $m_B = 1150$ MeV
is an average baryon mass.

One sees that
if the small effects of $\zeta$ are neglected, the three measurable ratios given in
Eqs. (26)-(28) depend in the $SU(5)$ model ($\kappa = 0$) on no unknown model parameters,
giving three testable predictions. In the $SO(10)$ model they depend on only the one unknown model parameter $\kappa$, giving two testable predictions. 
The $SO(10)$ version is more interesting, however, in that the value of $b$ can be extracted in two independent ways from these two predictions and compared to the 
value of $b$ obtained from fitting the CKM mixing, which is $b
= \frac{m_s}{m_d} V_{us}$. (See Eq. (13).) This is a highly non-trivial test of the model. And, indeed, this is one of the few models (maybe the only one) where proton decay allows a direct test of a model of quark and lepton masses and the mechanism of flavor mixing. 

Let us now turn to the operators involving $W_{(-1/3)}$, which give the nucleon decay modes with neutrinos.

\begin{equation}
\begin{array}{cl}
& e^{i ( \Phi_u - \Phi_d)_1} [\overline{\tilde{u}^c} \slashed{W}_{(-2/3)} D_1 \sum_{J=1,2,3} (A_{\Delta})_{1J} \nu_J]
\\ & \\
& + \;\; e^{i(\Phi_d + 2 \Phi)_1} [\overline{[c_{\theta} \tilde{d} + s_{\theta} \tilde{s}]} \slashed{W}_{(-2/3)} D_1 \sum_{J= 1,2}(A_{\Delta})_{1J} \tilde{d}^c_J] 
\\ & \\ & + \;\; 
e^{i(\Phi_d + 2 \Phi)_2} [\overline{[-s_{\theta} \tilde{d} + c_{\theta} \tilde{s}]} \slashed{W}_{(-2/3)} D_2 \sum_{J= 1,2}(A_{\Delta})_{2J} \tilde{d}^c_J] \\ & \\
= & e^{i ( \Phi_u - \Phi_d)_1}[ \overline{\tilde{u}^c} \slashed{W}_{(-2/3)} D_1 (\nu_e + b \nu_{\mu} + ce^{-i \theta} \nu_{\tau})] \\ & \\
 & + \;\; e^{i(\Phi_d + 2 \Phi)_1}  [\overline{[c_{\theta} \tilde{d} + s_{\theta} \tilde{s}]} \slashed{W}_{(-2/3)} D_1 (\tilde{d}^c + b \tilde{s}^c)] \\ & \\ 
 & + \;\; 
e^{i(\Phi_d + 2 \Phi)_2}  [\overline{[-s_{\theta} \tilde{d} + c_{\theta} \tilde{s}]} \slashed{W}_{(-2/3)} D_2 \tilde{s}^c] \\ & \\
\longrightarrow & 
e^{i ( \Phi_u - \Phi_d)_1} D_1 [\overline{\tilde{u}^c} \slashed{W}_{(-2/3)} (\nu_e + b \nu_{\mu} + ce^{-i \theta} \nu_{\tau})]
\\ & \\
& + \;\;  e^{i(\Phi_d + 2 \Phi)_1} D_1 [ c_{\theta} \overline{\tilde{d}} \slashed{W}_{(-2/3)} \tilde{d}^c 
+ (s_{\theta} + \lambda c_{\theta}) \overline{\tilde{s}} \slashed{W}_{(-2/3)} \tilde{d}^c
\\ & \\ & + \;\; (c_{\theta} b- \lambda s_{\theta})  \overline{\tilde{d}} \slashed{W}_{(-2/3)} \tilde{s}^c
+ (s_{\theta} b) \overline{\tilde{s}} \slashed{W}_{(-2/3)} \tilde{s}^c ],
\end{array}
\end{equation}

\noindent where $\lambda$ is a complex number with magnitude $D_2/D_1$, and phase
$(\Phi_d + 2 \Phi)_2 - (\Phi_d + 2 \Phi)_1$. Both the magnitude and phase of $\lambda$ are free parameters in this model, even if the mechanism of \cite{NelsonBarr} is implemented.

The terms in Eq. (30) allow us to write down the $d=6$ nucleon decay operators that come from the exchange of $W_{(-2/3)}$:

\begin{equation}
e^{i(- \Phi_u + 2 \Phi_d + 2 \Phi)_1} \kappa \left( \frac{g_{5}^2}{2M_{5}^2} \right)  \left[  
c_{\theta} \overline{\tilde{d}} \gamma_{\mu} \tilde{d}^c 
+ (s_{\theta} + \lambda c_{\theta}) \overline{\tilde{s}} \gamma_{\mu} \tilde{d}^c
+ (c_{\theta} b- \lambda s_{\theta})  \overline{\tilde{d}} \gamma_{\mu} \tilde{s}^c
+ (s_{\theta} b) \overline{\tilde{s}} \gamma_{\mu} \tilde{s}^c \right] \left[
\overline{\nu_*} \gamma^{\mu} \tilde{u}^c \right], 
\end{equation}

\noindent where $\nu_* \equiv (\nu_e + b \nu_{\mu} + ce^{-i \theta} \nu_{\tau})$.
This can be compared to the operator containing the neutrino fields coming from the exchange of $SU(5)$ gauge bosons, given in Eq. (19), which after Fierzing is

\begin{equation}
\left( \frac{g_5^2}{2M_5^2} \right)  
\left[  
\left(
c_{\theta} \overline{\tilde{d}} \gamma_{\mu} \tilde{d}^c 
+ s_{\theta} \overline{\tilde{s}} \gamma_{\mu} \tilde{d}^c
\right) (\overline{\nu_e}  \gamma^{\mu} \tilde{u}^c )
+ \left( 
c_{\theta}  \; \overline{\tilde{d}} \gamma_{\mu} \tilde{s}^c
+ s_{\theta} \;  \overline{\tilde{s}} \gamma_{\mu} \tilde{s}^c 
\right) 
(\overline{\nu_{\mu}} \gamma^{\mu} \tilde{u}^c ) 
\right]. 
\end{equation}

There are two measurable two-body decays of the proton that involve neutrinos, $p \rightarrow \pi^+ \nu$ and $p \rightarrow K^+ \nu$, since the flavor of the neutrino is not observable in practice. Thus, two more ratios of rates can be measured. In the $SU(5)$ version of the model, these do not depend on any additional model parameters, so two  predictions result for the neutrino modes. In the $SO(10)$ version of the model, an additional complex model parameter enters, namely $\lambda$. For some ranges of 
$|\lambda|$, the phase of $\lambda$ doesn't make much difference, so there would be one prediction for the neutrino modes. One can see form the definition of $\nu_*$ that this prediction would test the values of the parameters $a$ and $c$ that are given in Eq. (13).

To sum up, in the $SU(5)$ version of the model, there are five measurable ratios of rates for the two-body proton decays, and so there are five predictions if the effects of
the small parameter $\zeta$ can be neglected. These five predictions do not test the values of the quantities $a$, $b$, and $c$. On the other hand, if the effects of $\zeta$ could be measured with enough precision, the values of the phases $\alpha_1$ and $\alpha_2$ could be determined. If they are consistent with 0 or $\pi$, then it would support the mechanism for solving the Strong CP Problem proposed in \cite{NelsonBarr}. In the $SO(10)$ version of the model, the same five measurable ratios of proton decay rates depend (if $\zeta$ is neglected) on two unknown model parameters, $\kappa$ and $\lambda$. If the value of the latter is such that its phase does
not matter, then there are three predictions. These do test the quantities $a$, $b$, and $c$, and thus would provide a highly significant discriminant between this model and others. 

\section*{Acknowledgements} The authors acknowledge useful discussions
with the participants of the CETUP2012 and PHENO2013 workshops. This work was
supported by U.S. DOE under contract DE-FG02-12ER41808.

\end{document}